\documentclass[12pt]{article}
\usepackage{amssymb,amsmath,epsfig}

\begin{document}

\title{\bf  Cylindrically Symmetric Models of Anisotropic Compact Stars}
\author{ G. Abbas
\thanks{ghulamabbas@ciitsahiwal.edu.pk}, Sumara Nazeer
\thanks{sumara.nazeer@yahoo.com} and M. A. Meraj
\thanks{asad@ciitsahiwal.edu.pk}
\\Department of Mathematics, COMSATS Institute\\
of Information Technology, Sahiwal-57000, Pakistan.}
\date{}
\maketitle
\begin{abstract}
In this paper we have discussed the possibility of forming
anisotropic compact stars from cosmological constant as one of the
competent candidates of dark energy with cylindrical symmetry. For
this purpose, we have applied the analytical solution of Krori and
Barua metric to a particular cylindrically symmetric spacetime. The
unknown constants in Krori and Barua metric have been determined by
using masses and radii of class of compact stars like 4$U$1820-30,
Her X-1, SAX J 1808-3658. The properties of these stars have been
analyzed in detail. In this setting the cosmological constant has
been taken as a variable which depends on the radial coordinates. We
have checked all the regularity conditions, stability and surface
redshift of the compact stars 4$U$1820-30, Her
X-1, SAX J 1808-3658. \\
\end{abstract}

\section{Introduction}

The spherically symmetric static exterior and interior Schwarzschild
solutions of the Einstein field equations are well-known staples in
the elementary courses of General Relativity (GR). On the other hand
cylindrically symmetric static solutions (combining the translation
along the axis and rotation around the axis) are less familiar among
the general relativists. The static spherically vacuum solutions
were found in early 20th century by Weyl (1917) and Levi-Civita
(1919). They were interested in the more general problem of static
geometries that are axially symmetric. In this paper, we construct
and study cylindrically symmetric solutions of the anisotropic
strange stars that the best candidates of X-rays bluster. For this
purpose, we established the nonlinear differential equations for
cylindrically anisotropic source in the presence of varying
cosmological constant and find some analytic solutions which have
been applied to a class of strange stars.

Recent observational data and theoretical results in modern
cosmology revealed the facts that dark energy might be described in
a scientific way by the cosmological constant. The measurements
obtained by the Wilkinson Microwave Anisotropic Probe (WMAP) imply
that three-fourth of total mass-energy in our universe is dark
energy (DE)(Perlmutter et al. 1997, 1998, 1999 and Riess et al.
1998). The well-known leading theory of DE is mainly based on the
cosmological constant. characterized by the repulsive pressure as
defined by Einstein in 1917 for the formulation of static universe
model. Later on, Zel'dovich (1967) interpreted cosmological constant
as a vacuum energy of quantum fluctuation which has a size of order
of $\sim 3\times10^{-56}cm^{-2}$ (Peebles and Ratra 2003). Recent
astronomical observations like the study of type Ia supernova show
that the expansion of the universe is accelerating and it is
believed that this may be due to a non-vanishing positive
cosmological constant. These observations have attracted attention
to the study of astrophysical objects with cosmological constant.

In order to model the mass and radius of the Neutron star, Egeland
(2007) predicted that the existence of cosmological constant depends
upon the density of the vacuum. For this purpose Egeland used the
Fermi equation of state with the relativistic equation of
hydrostatic equilibrium. Motivated by this fact, we introduced the
cosmological constant in a small scale to study the structure of
strange stars and concluded that cosmological constant can describes
the class of some strange stars for example X-ray bluster
4$U$1820-30 X-ray pulsar Her X-1, Millisecond pulsar SAX J 1808-3658
etc., in very good manners. Dey et al. (1998), Usov (1998), Ruderman
(2004), Mak and Harko (2002, 2003) and many others have studied the
structure of strange stars by using different approaches.

 Mak and Harko (2004) presented a class of exact solutions of the
 field equation by using spherical symmetry. They also found that energy
 density as well as tangential and radial pressure are finite and
 positive inside the anisotropic star. Chaisi and Maharaj (2005) established
 an algorithm with an anisotropic matter distribution.
 By using Chaplygin gas equation of state (EoS), Rahaman et al. (2012) extended the Krori-Barua
 (1975) analysis of the charge anisotropic static spherically symmetric
 spacetime. Lobo (2006) generated the anisotropic exact models with a
 barotropic EOS for compact objects. He also generalized the
 Mazur-Mottola gravastar models by considering a matching of an
 interior solution governed by the dark energy EOS
  $\omega=\frac{p}{\rho}<\frac{-1}{3}$  to an exterior Schwarzschild
 vacuum solution at a junction interface. In this paper, we have formulated the cylindrically symmetric models
of strange stars which have been proposed earlier by Alcok et al.
(1986) and Haensel et al.(1986). Herrera and his collaborators
(Herrere 1992, Herrera et al. 2008a, 2008b, 2011) have discussed the
stability and gravitational collapse of anisotropic stars. The
related work in modified theory of gravity has also been done by
Sharif and Abbas (2013a, 2013b, 2013c). Hossein et al. (2012) have
studied the properties of the anisotropic compact stars in the
presence of cosmological constant.

The cylindrically symmetric models proposed here are associated with
cosmological constant and we have studied the stability of the model
by calculating the speed of sound using the anisotropic property of
the model. Finally, the surface red shift has been calculated using
the observational data of a class of anisotropic stars. The plan of
the paper is following. In the next section, we present the
anisotropic source and Einstein field equations. Section \textbf{3}
deals with the physical analysis of the proposed model. In the last
section, the results of the paper are concluded.

\section{Interior Matter Distribution and Field Equations}

For solving Einstein equations with a cylindrically symmetric
spacetime, firstly, we must assume the line element with which we
would like to work. Let $t$ denote a time coordinate, for a fixed
time $t$, a cylindrically symmetric spacetime can be described as
follows. There is a central axis of symmetry, with $z$ denoting the
coordinate along the static solutions of Einstein's equations. The
general static cylindrically symmetric spacetime (Brito et al. 2012)
is given by
 \begin{equation}\label{1}
 ds^2=-e^{\nu(r)}dt^2+e^{\mu(r)}dr^2+e^{\alpha(r)}dz^2+e^{\beta(r)}d{\theta^2},
\end{equation}
where $\nu,~~\mu,~~\alpha$ and $\beta$ are unknown functions. In
analogy to standard spherically symmetric spacetime, we define
coordinate $r$ in such a way that co-efficient of $d{\theta^2}$ is
equal to $r^2$. This transformation is called tangential gauge, thus
by setting $e^{\beta(r)}=r^2$, metric (\ref{1}) can be written as
\begin{equation}\label{2}
 ds^2=-e^{\nu(r)}dt^2+e^{\mu(r)}dr^2+e^{\alpha(r)}dz^2+r^2d{\theta^2}.
\end{equation}
Since co-efficient of $dr^2$ and $dz^2$ have same dimensions, thus
it is convenient to take $e^{\mu(r)}=e^{\alpha(r)}$. Hence metric
(\ref{2}) reduces to
 \begin{equation}\label{3}
 ds^2=-e^{\nu(r)}dt^2+e^{\mu(r)}(dr^2+dz^2)+r^2d{\theta^2}.
\end{equation}
 In this equation $\mu(r)=Ar^2$ and $\nu(r)=Br^2+C$ (Krori and Barura 2012)
 where $A, B$ and $C$ are arbitrary constants to be determined by using some boundary conditions.
The interior of compact object may be defined in terms of
anisotropic fluid which has following form
\begin{equation}\label{4}
  T_{\alpha\beta}=(\rho+P_t) U_{\alpha} U_{\beta}+P_{t}
  g_{\alpha\beta}+(P_r-P_t)\psi_{\alpha}\psi_{\beta},
\end{equation}
where $U_\alpha=e^{\frac{\nu}{2}}
\delta_\alpha^0,~~~\psi_\alpha=e^{\frac{\mu}{2}} \delta_\alpha^1$,
$\rho$,  $P_t$ and $P_r$ correspond to the energy density,
transverse and radial pressures, respectively. In this case
cosmological constant has radial dependence such that
$\Lambda=\Lambda(r)=\Lambda_r$.
 Therefore, the Einstein field equations
\begin{equation}\label{5}
G_{\alpha \beta}= R_{\alpha\beta}-\frac{1}{2}g_{\alpha\beta}R +
 g_{\alpha\beta}\Lambda \equiv \frac{{8\pi G}}{c^4}T_{\alpha\beta},
\end{equation}
 for the metric in Eq.(\ref{3}) (in the relativistic units $G=c=1$) are obtained as
 follows:
\begin{eqnarray}\label{6}
 8\pi\rho+{\Lambda}_r&=&-\frac{1}{2}\left(\frac{\mu''}{e^{\mu(r)}}\right),\\\label{7}
 8\pi
 P_r-{\Lambda}_r&=&\frac{1}{4e^{\mu}}\left(\frac{\nu'^2(r)r+2r\nu''(r)+4\nu'(r)}{r}\right),\\\label{8}
 8\pi
 P_t-{\Lambda}_r&=&\frac{1}{4}\left[\frac{\nu'^2(r)+2\mu''(r)+2\nu''(r)}{e^{\mu(r)}}\right].
\end{eqnarray}

We assume that radial pressure of the compact star is proportional
to the matter density, so
\begin{equation}\label{9}
 P_r=m\rho,~~~~m>0,
 \end{equation}
where $m$ is the equation of state parameter.
 Now, from the metric (\ref{3}) and Eqs.(\ref{6})-(\ref{8}), we get the
 energy density $\rho$, tangential pressure $P_t$, radial pressure $P_r$ and cosmological parameter
 $\Lambda_r$. These quantities are
\begin{eqnarray}\label{10}
\rho&=&\frac{(12B-A+4B^2r^2)e^{-Ar^2}}{8\pi(m+1)},\\\nonumber
\\\label{11}
 P_r&=&\frac{e^{-Ar^2}[(m-7)(4B^2r^3+12Br)-8Arm]}{64\pi r (m+1)},
\\\label{12}
P_t &=&\frac{e^{-Ar^2}[B^2r^2(m-3)+B(m-11)+A]}{8\pi(m+1)},
\\\label{13}
{\Lambda}_r&=&\frac{e^{-Ar^2}}{m+1}[Am-4B(Br^2+3)].
\end{eqnarray}
The equation of state (EoS) parameters corresponding to normal and
transverse directions can be written as,
 $$P_r=\omega_r\rho,$$ then
from Eqs.(\ref{9}), we get
\begin{equation}\label{14}
\omega_r(r)=m.
\end{equation}
Also, when  $$P_t=\omega_t\rho,$$ then from Eqs.(\ref{10}) and
(\ref{12}), we get
\begin{equation}\label{15}
\omega_t(r)=\frac{B^2r^2(m-3)+B(m-11)+A}{12B-A+4B^2r^2}.
\end{equation}
\section{Physical Analysis}
In this section, we shall discuss following features of our model:
\subsection{Anisotropic Behavior}
Taking derivative of Eqs. (\ref{10}) and (\ref{11}), we get
\begin{eqnarray}\label{16}
\frac{d\rho}{dr}&=&-\left[\frac{(24BAr-2A^2r+8AB^2r^3)e^{-Ar^2}}{8\pi(m+1)}-\frac{8B^2re^{-Ar^2}}{8\pi(m+1)}\right]<0,
\\\label{17}
\frac{dP_r}{dr}&=&-\left[\frac{(24BAr-B^2r(1+Ar^2))e^{-Ar^2}}{8\pi(m+1)}-\frac{A^2mre^{-Ar^2}}{4\pi(m+1)}\right]<0.
\end{eqnarray}
At center $r=0$, our model provides that
\begin{eqnarray}\label{18}
\frac{d\rho}{dr}&=&0,~~~ \frac{dP_r}{dr}=0\\\label{20}
\frac{d^2\rho}{dr^2}&<&0,~~~~
 \frac{d^2P_r}{dr^2}<0
\end{eqnarray}
which indicate maximality of radial pressure and density. This
implies the fact that $\rho$ and $P_r$ are decreasing function of
$r$ as shown in figures \textbf{1-6} for a class of strange star.
Similar behavior of $P_t$ and ${\Lambda}_r$ is shown in figure
\textbf{7-9} and \textbf{10-12}. The measure of anisotropy is
$$\Delta=\frac{2}{r}(P_t-Pr),$$ which takes the form
\begin{equation}\label{21}
 \Delta=\frac{1}{16\pi r
 }\left[e^{-Ar^2}\left(B(Br^2+1)+2A\right)\right].
\end{equation}
The anisotropy will be directed outward when $P_t>P_r$ this implies
that $\Delta>0$ and directed inward when $P_t<P_r$ implying
$\Delta>0$. In this case $\Delta>0$, for larger value of $r$ for a
class of strange stars as shown in figures \textbf{13-15}. This
implies that anisotropic force allows the construction of more
massive star while near the center there is attractive force as
$\Delta<0$ in figures \textbf{14, 15}. Note that the bound on the
EoS parameter $0<\omega_t(r)<1$ is shown in figure \textbf{16}. This
shows that star consists of ordinary matter and effect of
cosmological constant $\Lambda$.

\subsection{Matching Conditions}

Here, we match the interior metric (\ref{3}) to the vacuum exterior
cylindrically symmetric metric (Lemos and Zanchin 1996) given by
\begin{equation}\label{21}
 ds^2=-\left({\Lambda r^2}-\frac{4M}{r}\right)dt^2+\left({{\Lambda} r^2}-
 \frac{4M}{r}\right)^{-1}dr^2+{\Lambda}r^2dz^2+r^2d
 \theta ^2,
\end{equation}
where ${\Lambda}<0$ is cosmological constant.
  At the boundary $r=R$ continuity of the metric functions $g_{tt}$,
  $g_{rr}$ and $\frac{\partial g_{tt}}{\partial r}$ at the boundary
  surface yield,
\begin{eqnarray}\label{22}
  g_{tt}^-=g_{tt}^+,~~~~~
   g_{rr}^-=g_{rr}^+,~~~~~
   \frac{\partial g_{tt}^-}{\partial r}=\frac{\partial g_{tt}^+}{\partial r},
  \end{eqnarray}
where $-$ and $+$, correspond to interior and exterior solutions.
From the interior and exterior matrices, we get
 \begin{eqnarray}\label{23}
  A&=&\frac{1}{R^2}ln\left(\frac{4M}{
  R}+\sqrt{\frac{16M^2}{R^2}+4}\right),\\\label{24}
 B&=&\frac{1}{R^2}\left[\frac{\Lambda R^3+2M}{\Lambda R^3-4M}\right].
\end{eqnarray}
For the given values of $M$ and $R$ for given star, the constants
$A$ and $B$ are given in the table \textbf{1}.

\begin{figure}
\center\epsfig{file=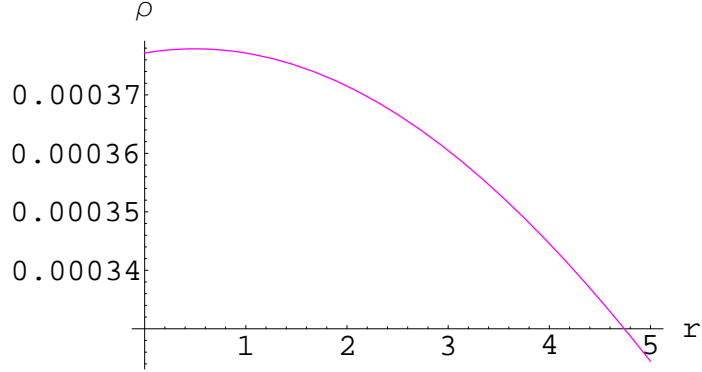, width=0.7\linewidth} \caption{Density
variation of Strange star candidate 4U 1820 - 30}
\end{figure}
\begin{figure}
\center\epsfig{file=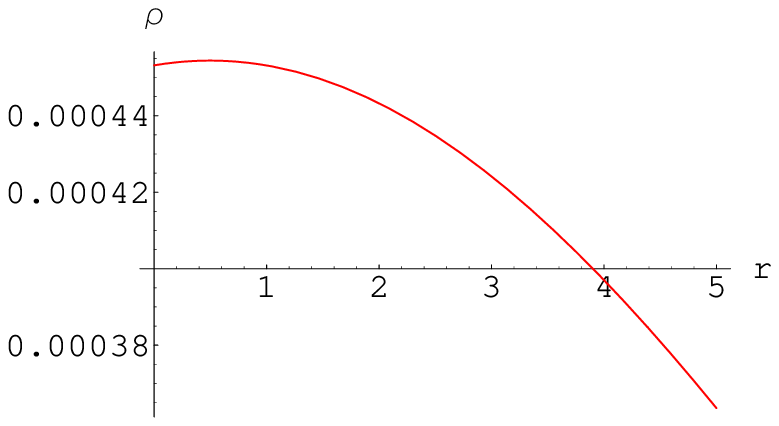, width=0.7\linewidth} \caption{Density
variation of Strange star candidate Her X - 1.}
 \center\epsfig{file=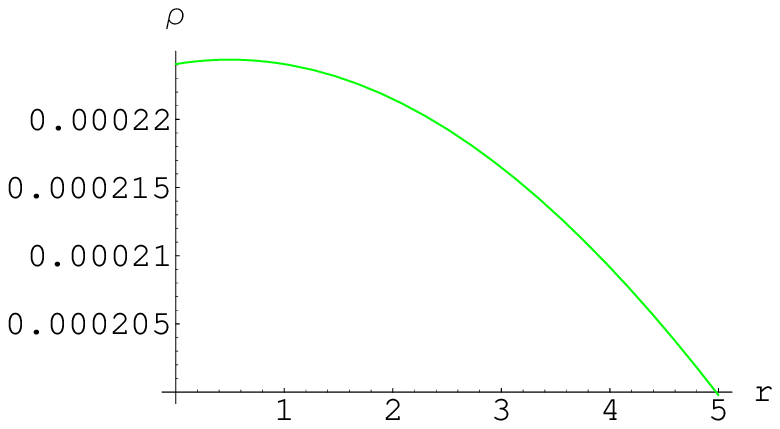, width=0.7\linewidth}
\caption{Density variation of Strange star candidate SAX J
1808.4-3658(SS1).}
\end{figure}
\begin{figure}
\center\epsfig{file=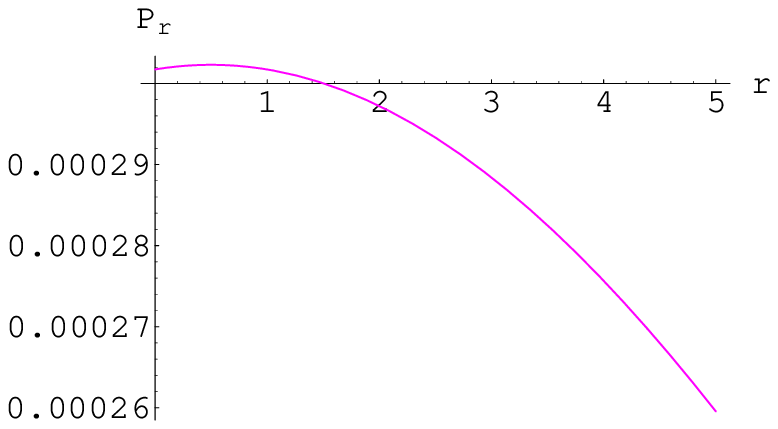, width=0.7\linewidth} \caption{Radial
pressure variation of Strange star candidate 4U 1820 - 30}
 \center\epsfig{file=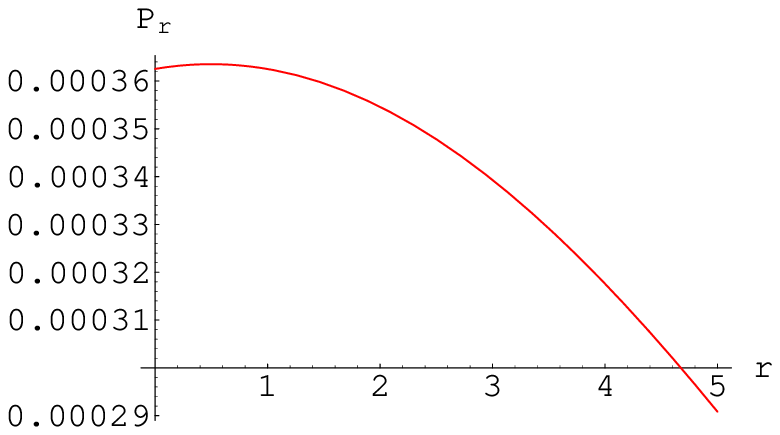, width=0.7\linewidth}
\caption{Radial pressure variation of strange star candidate Her X -
1.}
\end{figure}
\begin{figure}
\center\epsfig{file=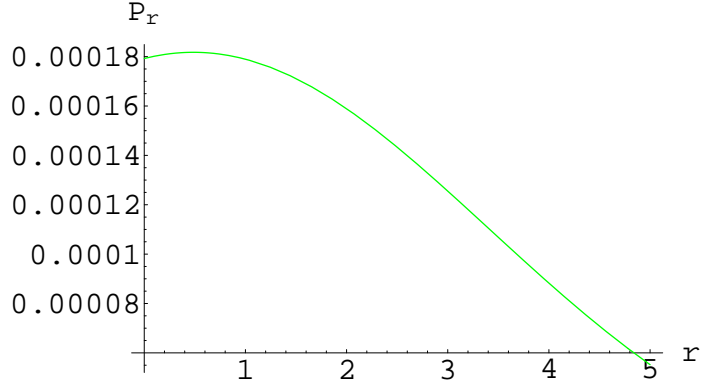, width=0.7\linewidth} \caption{Radial
pressure variation of strange star candidate SAX J
1808.4-3658(SS1).}
\end{figure}
\begin{figure}
\center\epsfig{file=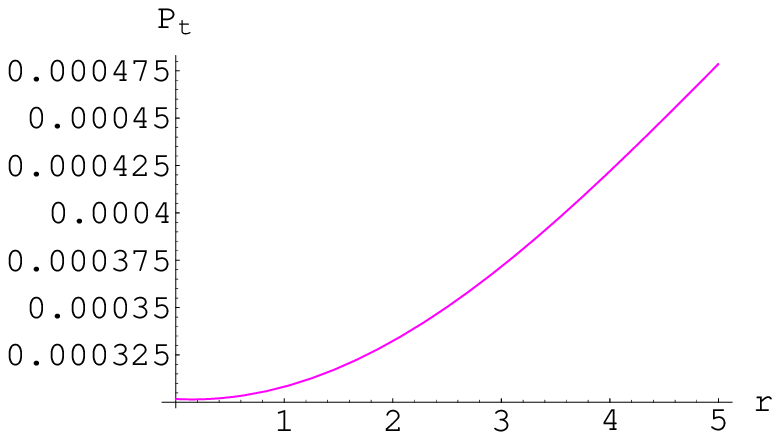, width=0.45\linewidth}
\caption{Tangential pressure variation of Strange star candidate 4U
1820 - 30.}
 \center\epsfig{file=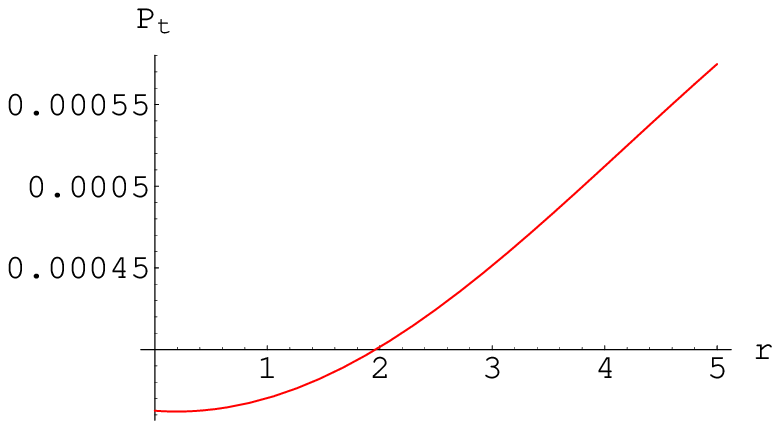, width=0.7\linewidth}
\caption{Tangential pressure variation of strange star candidate Her
X - 1.}
\end{figure}
\begin{figure}
 \center\epsfig{file=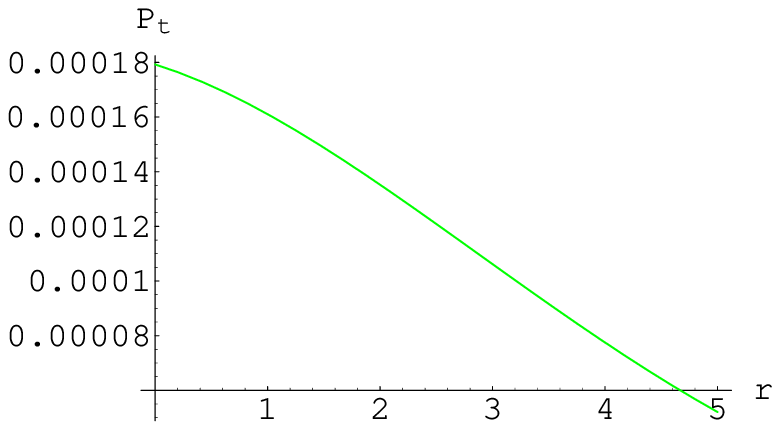, width=0.7\linewidth}
\caption{{Tangential pressure variation of strange star candidate
SAX J 1808.4-3658(SS1).}}
 \center\epsfig{file=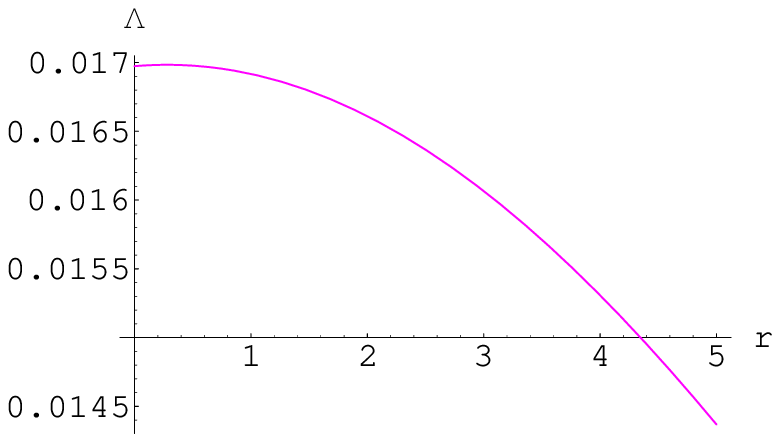, width=0.7\linewidth}
\caption{Behavior of cosmological constant for strange star
candidate 4U 1820 - 30.}
\end{figure}
\begin{figure}
 \center\epsfig{file=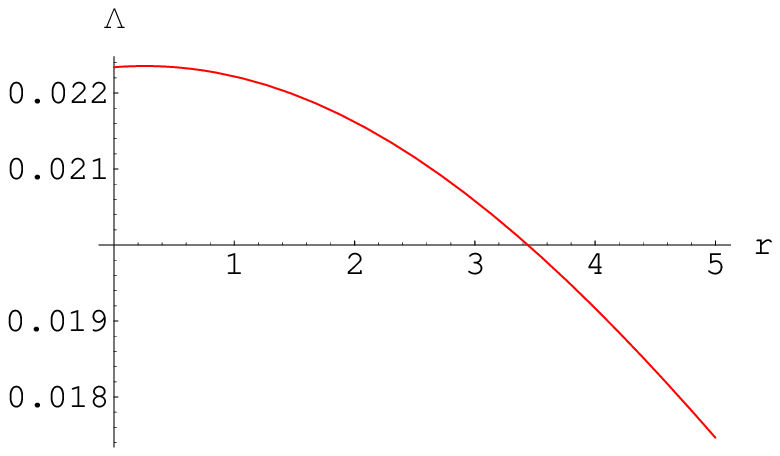, width=0.7\linewidth}
\caption{Behavior of cosmological constant for strange star
candidate Her X - 1.} \center\epsfig{file=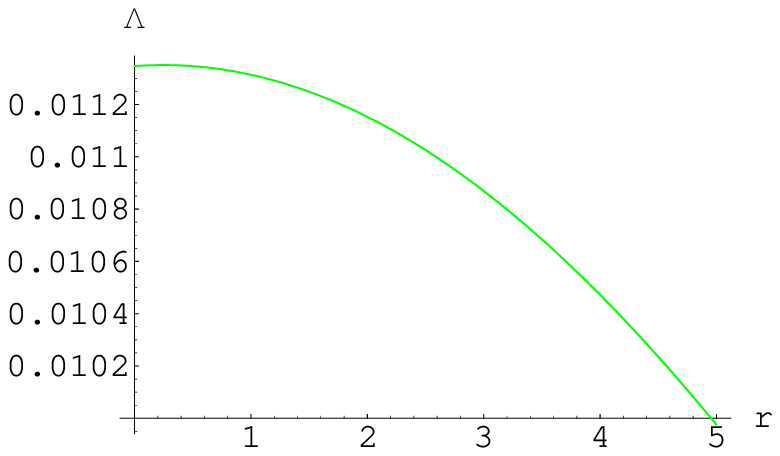,
width=0.7\linewidth} \caption{Behavior of cosmological constant for
strange star candidate SAX J 1808.4-3658(SS1).}
\end{figure}
\begin{figure}
\center\epsfig{file=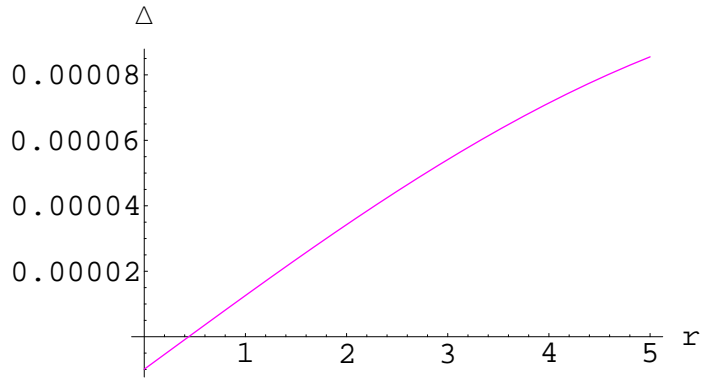, width=0.7\linewidth}
\caption{Anisotropic behavior of strange star candidate 4U 1820 -
30.}
\end{figure}
\begin{figure}
\center\epsfig{file=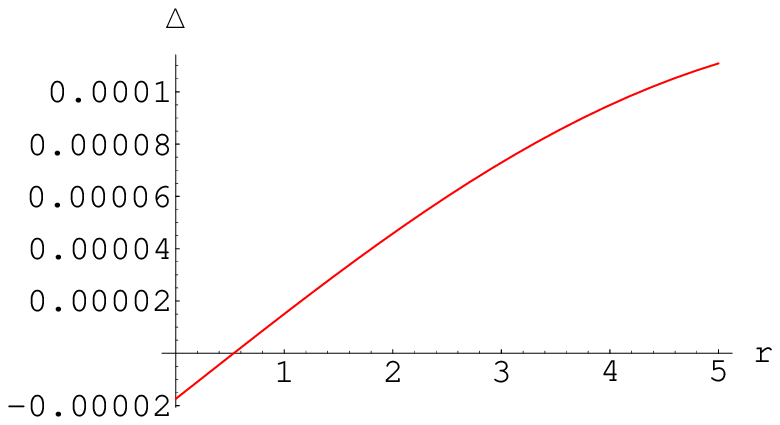, width=0.7\linewidth}
\caption{Anisotropic behavior of strange star candidate Her X - 1.}
 \center\epsfig{file=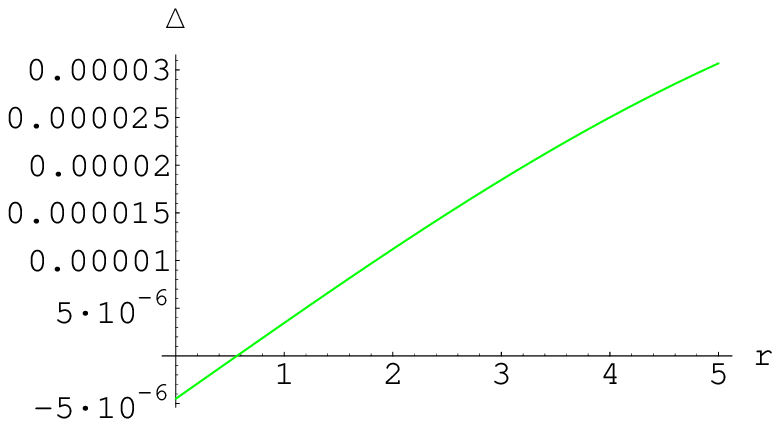, width=0.7\linewidth}
\caption{Anisotropic behavior of strange star candidate SAX J
1808.4-3658(SS1).}
\end{figure}
\begin{figure}
 \center\epsfig{file=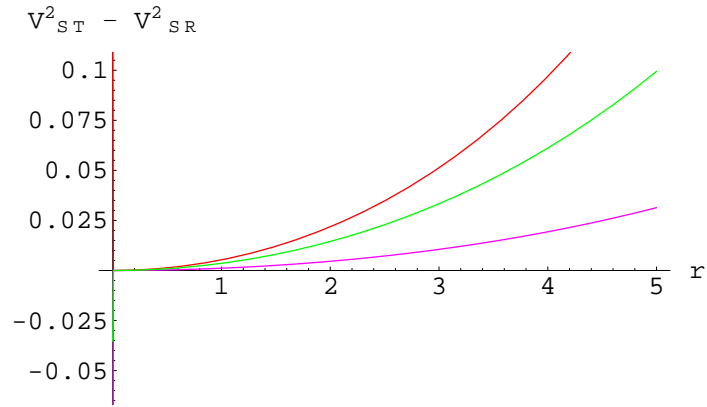,
width=0.7\linewidth} \caption{Variation of ${v^2}_{ST} - {v^2}_{SR}$
of Strange star candidate $4U 1820$ - 30, $Her X - 1$ and SAX J
1808.4-3658(SS1)}
\end{figure}
\begin{figure}
 \center\epsfig{file=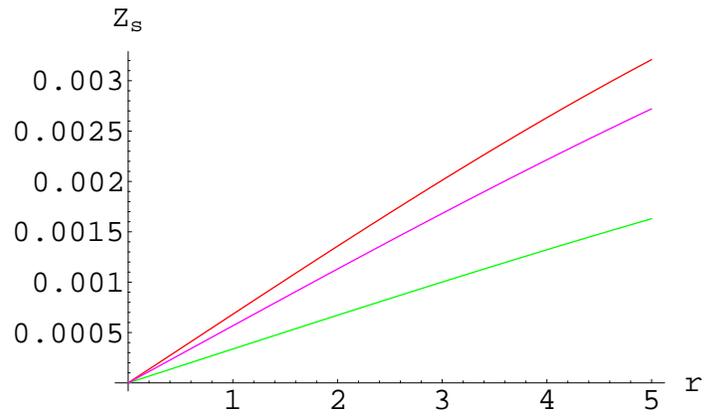, width=0.7\linewidth}
\caption{Behavior of redshift of Strange star candidate  $4U 1820$ -
30, $Her X - 1$ and SAX J 1808.4-3658(SS1).}
\end{figure}

\subsection{Stability}
We define sound speed as,
\begin{eqnarray}\label{43}
\upsilon^2_{SR}&=&\frac{dP_r}{d \rho}=
\left[\frac{4BAr-B^2r(1+Ar^2)-2A^2mr}{24BAr-2A^2r+8AB^2r^3-8B^2r}\right]\\
\upsilon
^2_{ST}&=&\frac{dP_t}{d \rho}\nonumber\\
&=&\left[\frac{2B^2r(m-3)-2Ar(B^2r^2(m-3)+B(m-11)+A)}{24BAr-2A^2r+8AB^2r^3-8B^2r}\right].\nonumber\\
\end{eqnarray}
\begin{table}[ht]
\caption{Values of constant for given Masses and Radii of Stars}
\begin{center}
\begin{tabular}{|c|c|c|c|c|c|}
\hline {Strange Quark Star}&  \textbf{ $M$} & \textbf{$R(km)$} &
\textbf{ $\frac{M}{R}$} &\textbf{ $A(km ^{-2})$}& \textbf{$B(km
^{-2})$}
\\\hline  Her X-1& 0.88$M_\odot$& 7.7&0.168&0.00749431669  &
$0.017062831$
\\\hline SAX J 1808.4-3658& 1.435$M_\odot$& 7.07&0.299& 0.010949753 &
$0.020501511$
\\\hline 4U 1820-30&2.25$M_\odot$& 10.0 &0.332&0.005715628647 &
$0.0101366226$
\\\hline
\end{tabular}
\end{center}
\end{table}
These quantities are shown in figure \textbf{17}, in this figure
$I=R,~S$. From the above equations, we get
\begin{eqnarray}\nonumber
\upsilon^2_{ST}-\upsilon^2_{SR}&=&\frac{2B^2rm-5B^2r-2B^2Ar^3m+5B^2Ar^3+18ABr-2ABrm-2A^2r+2A^2rm}{24BAr-2A^2r+8AB^2r^3-8B^2r},\nonumber\\
\end{eqnarray}
which can be simplified to the following form
\begin{eqnarray}
\upsilon^2_{ST}-\upsilon^2_{SR}&=&\frac{B^2(2rm-5r-2Ar^3m+5Ar^3)+2B(9Ar-2Arm)-2A^2r(1-m}{24BAr-2A^2r+8AB^2r^3-8B^2r}.\nonumber\\
\end{eqnarray}
From figure \textbf{{17}}, we can see that
$\mid\upsilon^2_{st}-\upsilon^2_{sr}\mid\leq1$. This is used to
check whether local anisotropic matter distribution is stable or
not. For this, we use the cracking concept (Herrere 1997) which
explain that potentially stable region is that region where radial
speed of sound is greater than the transverse speed of sound. Hence,
our proposed compact star model is stable.

\subsection{Surface Redshift}
The compactness of the star is given by
\begin{equation}
u=\frac{M}{b}=\frac{1}{2(m+1)}\left[\frac{2B^2e^{-Ar^2}}{A^2}-\frac{\sqrt{\pi}Erf[\sqrt{A}b](2B^2-A(A-12B))}{2bA^{\frac{5}{2}}}\right],
\end{equation}
where $b=r$. The surface redshift $(Z_s)$ corresponding to the above
compactness $(u)$ is obtained
\begin{equation}
1+{Z_s}=[1-2u]^{-\frac{1}{2}}=\left[1-\frac{1}{(m+1)}\left(\frac{2B^2e^{-Ar^2}}{A^2}-\frac{\sqrt{\pi}Erf[\sqrt{A}b](2B^2-A(A-12B))}{2bA^{\frac{5}{2}}}\right)\right]^{\frac{1}{2}}.
\end{equation}
The maximum surface redshift for the compact objects is given by
figure \textbf{18}.

\section{Conclusion}

In GR cylindrically symmetric  solution turns out to be analogous to
spherically symmetric solutions in many ways but also remain quite
different in many aspects. The first well-known static spherically
symmetric vacuum solution of Einstein Field equation is
Schwarzschild exterior solution characterized by mass, but there are
some static vacuum cylindrically symmetric solutions other than cone
or cosmic string solutions characterized by the defect angle. A cone
solution is Lorentz-invariant along the cylinder axis and hence
cannot arise from a matter source unless the matter source satisfies
the physical equation of state.

In this paper, we have constructed analytical solutions for the
compact stars with more general interior source and exterior
geometry. The analysis has been done by considering that stars are
anisotropic in their internal configuration. The present day
acceleration of the Universe in the form of DE allows us to consider
the cosmological constant as a variable in its character. The
interior configuration of the cylindrical star has been treated by
metric assumption. By the physical interpretation of the results, we
conclude that bound on effective EoS parameter is given by
$0<{\omega}_i<1$ which is in agreement with normal matter
distribution. The density and pressure attain the maximum value at
the center. It has been found that the anisotropy will be directed
outward when $P_t>P_r$ this implies that $\Delta>0$ and directed
inward when $P_t<P_r$ implying $\Delta>0$. In this case $\Delta>0$,
for larger value of $r$ for a class of strange stars as shown in
figures \textbf{13-15}. This implies that anisotropic force allows
the construction of more massive star while near the center of the
interior configuration there is attractive force as $\Delta<0$ in
figures \textbf{14, 15}. The range of $Z_s$ for the  compacts
objects  in this case lies in the range $0<Z_s\leq0.003$. In case of
isotropic interior configuration without cosmological constant this
range turnout to be $Z_s\leq2$. Hence in present configuration
redshift has been decreased to a certain range. According to
B$\ddot{o}$hmer and Harko anisotropic stars in the presence of
cosmological constant has the redshift value in the range
$Z_s\leq5$, which is consistent with the Ivanov (2002) bound
$Z_s\leq 5.211$. On the basis of the cracking concept, we have
discussed the stability of the proposed model and found that present
model is stable.

\vspace{0.25cm}

{\bf Acknowledgment}

\vspace{0.25cm}
 We highly appreciate the
fruitful comments of the anonymous referee for the improvements of
the paper.

\end{document}